\documentclass[conference]{IEEEtran}
\IEEEoverridecommandlockouts
\usepackage{cite}
\usepackage{amsmath,amssymb,amsfonts}
\usepackage{algorithmic}
\usepackage{graphicx}
\usepackage{textcomp}
\usepackage{xcolor}
\def\BibTeX{{\rm B\kern-.05em{\sc i\kern-.025em b}\kern-.08em
    T\kern-.1667em\lower.7ex\hbox{E}\kern-.125emX}}

\usepackage[english,spanish]{babel}

\usepackage[utf8]{inputenc}
\usepackage[T1]{fontenc}
\usepackage{color}
\usepackage{tabularx,booktabs}
\usepackage{graphicx}
\usepackage{lipsum}
\usepackage{url}
\usepackage{breakurl}
\usepackage[breaklinks]{hyperref}
\usepackage{array}
\usepackage{subfigure}

\begin{document}
\selectlanguage{english}

\title{Development of Computational Thinking in High School Students: A Case Study in Chile}

\author{\IEEEauthorblockN{Felipe González}
\IEEEauthorblockA{\textit{Departamento de Informática} \\
\textit{UTFSM}\\
Santiago, Chile \\
felipe.gonzalezpi@usm.cl}
\and
\IEEEauthorblockN{Claudia López}
\IEEEauthorblockA{\textit{Departamento de Informática} \\
\textit{UTFSM}\\
Valparaíso, Chile \\
claudia@inf.utfsm.cl }
\and
\IEEEauthorblockN{Carlos Castro}
\IEEEauthorblockA{\textit{Departamento de Informática} \\
\textit{UTFSM}\\
Valparaíso, Chile\\
carlos.castro@inf.utfsm.cl }
}
\IEEEoverridecommandlockouts
\IEEEpubid{\makebox[\columnwidth]{978-1-5386-9233-2/18/\$31.00~\copyright2018 IEEE \hfill} \hspace{\columnsep}\makebox[\columnwidth]{ }}
\maketitle
\IEEEpubidadjcol

\begin{abstract}
Most efforts to incorporate computational thinking in K-12 education use visual tools (e.g., Scratch) and focus on students in their first grades. Fewer projects investigate the development of computational thinking in students in the last years of school, who usually have not had early formal preparation to acquire these skills. This study provides evidence of the effectiveness of teaching C++ (a low-level programming language) to develop computational thinking in high school students in Chile. By applying a test before and after a voluntary C++ workshop, the results reveal a significant improvement in computational thinking after the workshop. However, we observe that there was a tendency to drop out of the workshop among students with lower levels of initial computational thinking. Besides, tenth-grade students obtained lower final scores than eleventh and twelfth-grade students. These results indicate that teaching a low-level programming language is useful, but it has high entry-barriers.

\end{abstract}

\begin{IEEEkeywords}
Computational Thinking Test; C++; Programming in High School. 
\end{IEEEkeywords}
\selectlanguage{spanish}

\section{Introducción}


Se estima que Chile para el año 2019 necesitará llenar un vacío de un 9\% de profesionales en el área de las Tecnologías de la Información y Comunicación. Además el país ha sido rápido en adopción de tecnología y las compañías han mostrado un alto interés en invertir en tecnologías emergentes \cite{pineda_y_gonzalez}. Sin embargo, la mayoría de las y los jóvenes que hoy cursan la educación secundaria no han desarrollado aún las habilidades de manejo de tecnología consideradas necesarias según el Ministerio de Educación de Chile \cite{simce_que_evalua}. El Sistema de Medición de la Calidad de la Educación (SIMCE) en las Tecnologías de la Información y Comunicación (TIC), realizado en su última versión en 2013, mostró que ``solo un $1.8\% $de los estudiantes presentaba un nivel avanzando en el uso de las TIC para la realización de tareas relacionadas con el aprendizaje y el conocimiento''\cite{SIMCE_PAIS_DIGITAL}.  

Para acercar a los alumnos al mundo digital, el Ministerio de Educación de Chile (MINEDUC) ha implementado el programa: ``Enlaces, Centro de Educación y Tecnología del Ministerio de Educación de Chile'' (Enlaces), que tiene como misión mejorar la calidad de la educación integrando la informática educativa en el sistema escolar y ha sido un  promotor del uso de las TIC dentro del aula de clases. Esto involucra un apoyo a la mejora de la infraestructura tecnológica y una serie de capacitaciones a docentes. Sin embargo, enfocarse únicamente en el uso de las TIC podría provocar que los alumnos aprendan solamente a utilizar soluciones ya configuradas, perdiendo la oportunidad de innovar y de crear soluciones nuevas a problemas de la actualidad \cite{simmonds_y_cols}. 

Un enfoque menos considerado hasta ahora por el MINEDUC ha sido el desarrollo del pensamiento computacional, el cual se refiere a una habilidad que involucra la resolución de problemas, diseño de sistemas y la comprensión del comportamiento humano haciendo uso de conceptos fundamentales de la informática \cite{wing_2006}. El pensamiento computacional  permite fomentar el pensamiento crítico, la innovación y creatividad de quienes lo desarrollen \cite{simmonds_y_cols}.

El pensamiento computacional ha tomado fuerza en países alrededor del mundo tras la modificación de los currículums escolares \cite{modificacion_curriculum} y la incorporación de la programación como una herramienta para el desarrollo de esta habilidad \cite{summary_ct}. Contraria a esta tendencia, la programación aún no ha sido incluida dentro del currículum de enseñanza científico-humanista en Chile. Han sido las fundaciones y universidades del país las que han comenzado a explorar ese camino a través de talleres de programación. Aunque estas actividades han captado interés y participación, hasta la fecha hay escasa evidencia del impacto de este tipo de instancias en el desarrollo del pensamiento computacional de  escolares en Chile. 

Este trabajo reporta el diseño, ejecución y análisis de un estudio del desarrollo del pensamiento computacional de estudiantes de enseñanza media científico-humanista que participaron en un taller de 30 horas de programación en C++ realizado en la Universidad Técnica Federico Santa María (UTFSM) durante el 2017. Los resultados obtenidos permiten proveer evidencia empírica inicial del impacto que genera los talleres de programación en contextos pre-universitarios. 

\section{Desarrollo del pensamiento computacional en educación pre-universitaria}


El pensamiento computacional se refiere a los procesos involucrados en formular un problema y expresar su solución de manera que un agente que procesa información - humano o máquina - la pueda llevar a cabo efectivamente \cite{wing2014computational}. 
Está inspirado en las habilidades que se utilizan al generar una solución computacional \cite{wing_2006}, e incluye habilidades tales como: organizar y analizar datos de manera lógica, representar datos a través de abstracciones, automatizar soluciones usando pensamiento algorítmico, construir soluciones iterativamente depurando soluciones anteriores, entre otras \cite{barr2011computational}. El pensamiento computacional no reemplaza la creatividad, el razonamiento o el pensamiento crítico, sino que las fortalece haciendo que las personas sean capaces de visualizar nuevas formas de organizar y solucionar un problema haciendo uso de tecnología \cite{simmonds_y_cols}. Sus proponentes postulan que es una habilidad que toda persona debiera desarrollar en el futuro, y es importante desarrollarla en la educación pre-universitaria \cite{wing_2006,barr2011computational}.

El pensamiento computacional se puede conceptualizar a través de tres dimensiones \cite{brennan}:
\begin{itemize}
    \item Conceptos computacionales: Involucra la enseñanza de elementos de la programación como direcciones (secuencias), ciclos, funciones, estructuras condicionales y otros.
    \item Prácticas computacionales: Involucra experimentar y crear pruebas de las soluciones creadas, depurar, reutilizar y remezclar código junto con distinguir lo esencial de un problema.
    \item Perspectivas computacionales: Involucra expresar, cuestionar y llegar a acuerdos con otros individuos en el proceso de búsqueda y creación de soluciones algorítmicas. 
\end{itemize}

La literatura distingue dos categorías de mecanismos para desarrollar el pensamiento computacional \cite{kalelioglu2016framework}:
\begin{itemize}
    \item Actividades que no utilizan dispositivos digitales, tales como juegos lógicos, \textit{storytelling}, entre otros. 
    \item Actividades computarizadas que involucran distintas tareas de programación \cite{summary_ct}, ya sea utilizando herramientas visuales (como \textit{Scratch}  o \textit{Alice}) o  
    lenguajes de programación de bajo nivel (como \textit{Java} o \textit{C++}).
\end{itemize}

Un estudio con estudiantes de enseñanza básica evidenció que los mecanismos de la primera categoría son efectivos sólo hasta cierto nivel del pensamiento computacional, pero dejan de ser efectivos más allá de tal límite \cite{brackmann2017development}. Por otro lado, diversos estudios han reportado sobre la efectividad de los mecanismos de la segunda categoría para mejorar el pensamiento computacional de estudiantes desde cinco a los catorce años \cite{lye2014review}. El diseño de algoritmos involucrado en actividades de la segunda categoría no solo fortalece habilidades del ámbito del pensamiento computacional, sino que también aporta en la elaboración de estrategias de resolución de problemas potencialmente útiles en ámbitos no computacionales \cite{cornejo2017approach}.

Gran parte de la investigación que involucra tareas de programación para estudiantes escolares ha evidenciado la efectividad del uso de herramientas visuales de programación \cite{lye2014review}. Por ejemplo, una intervención de dos años en cinco escuelas españolas introdujo el uso de \textit{Scratch} en las asignaturas de ciencias y arte de 107 estudiantes de sexto año \cite{espana_experimento}. Los alumnos podían crear proyectos \textit{Scratch} acordes a las temáticas de su curso. La aplicación del test \textit{Visual Blocks Creative Computing} antes y después de la intervención, mostró que los estudiantes mejoraron su dominio respecto a conceptos y prácticas computacionales. Además, otro estudio en estudiantes españoles de sexto básico sin conocimientos previos de programación mostró que la realización de  proyectos en \textit{Scratch} puede generar progreso en conceptos computacionales específicos, tales como: ciclos, direcciones y funciones \cite{espana_experimento_tcp}. En este último caso, la medición del desarrollo del pensamiento computacional estuvo sujeta al test diseñado por \cite{TPC_roman} y fue utilizado junto a otros instrumentos de medición contenidos en la plataforma Dr. Scratch.  


\subsection{El contexto en Chile}
Mientras varios países están incluyendo la programación en los planes de estudio de educación primaria y secundaria como una forma de desarrollar el pensamiento computacional \cite{summary_ct}, Chile aún no concreta su incorporación en las bases curriculares de la educación escolar. Han sido organizaciones e investigadores quienes han diseñado iniciativas para fomentar el pensamiento computacional en estudiantes escolares en Chile. Entre ellas están:

%
%
%

\begin{itemize}
    \item ``Taller de Jóvenes Programadores’’, de BiblioRedes, que busca que toda persona mayor a ocho años residente en Chile pueda introducirse en el mundo de la programación aprendiendo, a través de módulos \textit{online}, algunas de las siguientes herramientas \cite{biblioredes}: \textit{Scratch}, \textit{Snap!}, \textit{AppInventor}, \textit{JavaScript}, \textit{CSS}, \textit{PHP} o \textit{Python}. 
    A la fecha, no se conocen públicamente evaluaciones de esta iniciativa.
    \item ``La hora del código Chile'', liderado por la fundación Kodea, intenta promover la enseñanza de programación en escolares invitándolos a seguir tutoriales interactivos en una actividad de una hora. En ellos, mediante programación visual se busca que los participantes fomenten habilidades de resolución de problemas, lógica y creatividad \cite{hour_of_code_chile}, \cite{hour_of_code_global}. Si bien no se han encontrado estudios sobre el impacto de esta actividad en el contexto nacional, en \cite{hour_of_code_impact} analizaron el impacto de esta actividad en estudiantes de los Estados Unidos. Un cuestionario pre-test y post-test arrojó que una vez finalizada la actividad los estudiantes adquirieron una actitud más positiva en relación a la programación. Sin embargo, sus habilidades en esta materia no se ven significativamente mejoradas por lo que los autores sugieren utilizar estos tutoriales interactivos en compañía de clases de programación realizadas de manera tradicional.

    \item ``Desarrollando el Pensamiento Computacional'', de la Corporación $C^{100}$, se enfoca en niños entre nueve y doce años sin experiencia previa de programación, quienes durante un taller de cinco días de duración deben desarrollar habilidades del pensamiento computacional mediante la realización de proyectos usando \textit{Scratch} \cite{taller_c100}. Un estudio asociado al taller en donde participaron 55 escolares sin experiencia previa de programación entre los 10 y 12 años,  reveló que los participantes son capaces de adquirir nociones iniciales de buenas prácticas de la ingeniería de software (tales como: revisión del propio código, entender requerimientos funcionales, entre otros), habilidades que promueven el pensamiento computacional. En la actualidad, se está investigando si el orden y la profundidad en que son vistos los contenidos del taller afecta el cómo asimilan los estudiantes distintos conceptos del pensamiento computacional al igual que las prácticas de la ingeniería de software asociadas \cite{simmonds_paper_k6}.


    \item ``Niñas Pro(Gramadoras)'', es un taller de programación competitiva, liderado por la Corporación $C^{100}$, dirigido a alumnas de enseñanza media. Tiene como fin preparar a las participantes para las Olimpiadas Chilenas de Informática (OCI) junto con fomentar el ingreso de mujeres a las carreras vinculadas a la tecnología \cite{c100}. Entre sus contenidos se encuentran: tipos de datos, condicionales, ciclos, funciones, entre otros \cite{ninas_programadoras_temario}. Hasta la fechan, no se han reportado estudios sobre el impacto de esta actividad. 
    \item Por otro lado, un estudio evaluó  la realización de actividades de una hora de duración usando \textit{Scratch} para un total de 27 alumnos entre sexto básico y segundo año medio distribuidos en las ciudades de Viña del Mar y Linares \cite{linares_talca_experimento}. Los resultados indicaron que se logró un avance en el desarrollo del pensamiento lógico y algorítmico una vez que se utilizó la herramienta. 
    
\end{itemize}

\subsection{Pregunta de investigación}

Tanto a nivel global como en el contexto nacional existe la tendencia a utilizar lenguajes de programación visuales, más que lenguajes de programación de bajo nivel, para desarrollar el pensamiento computacional a nivel escolar \cite{lye2014review}. Aunque menos poderosos, los lenguajes de programación visuales son considerados más fáciles de aprender ya que tienen una sintaxis reducida y, por lo tanto, implican una menor carga cognitiva para los aprendices \cite{lye2014review}. Al mismo tiempo, la mayoría de los estudios sobre educación pre-universitaria se enfocan en estudiantes de temprana edad, entre los cinco y catorce años.  

Considerando que existe un amplio número de estudiantes en el ciclo final de la enseñanza escolar y que no han sido involucrados en intervenciones de pensamiento computacional a edades tempranas en su educación formal, surge la pregunta que motiva nuestro trabajo: ¿qué mecanismos son efectivos para desarrollar el pensamiento computacional de jóvenes de enseñanza media?

Los lenguajes de programación de bajo nivel han sido ampliamente usado en universidades para entrenar a futuros profesionales de la computación, y existe evidencia de que aprender estos lenguajes fomenta el pensamiento computacional en estudiantes universitarios (ver resumen en \cite{lye2014review}). 

Dada esta brecha en que se prefiere que estudiantes de ciclos básicos usen lenguajes de programación visuales y estudiantes universitarios aprendan lenguajes de programación de bajo nivel, la pregunta que continúa abierta es ¿con qué mecanismo se puede desarrollar el pensamiento computacional de quienes están en los últimos años de la educación secundaria? O, puesto de otra manera por autores de un estudio en Chile, ¿qué  debe venir después del uso de \textit{Scratch}? \cite{linares_talca_experimento}. 

\section{Metodología}

Para comenzar a responder esta pregunta abierta, este estudio evalúa el desarrollo del pensamiento computacional en estudiantes de segundo a cuarto medio que asistieron voluntariamente a un taller de C++ organizado por una universidad chilena. El taller se sitúa en el contexto de preparación para la OCI, competencia orientada a estudiantes secundarios, que consiste en resolver problemas utilizando alguno de los siguientes lenguajes de programación: C, C++ o Pascal. La OCI busca difundir la ciencia de la computación y la informática a nivel nacional junto con descubrir tempranamente, alentar y reconocer a los jóvenes talentos en el área \cite{oci_descripcion}. 

Dado este contexto, el estudio se diseñó para contestar la siguiente pregunta de investigación: ¿Contribuye un taller de C++ al desarrollo del pensamiento computacional de estudiantes de enseñanza media de Chile?

\begin{figure}[!htpb]
\centerline{\includegraphics[width=\linewidth]{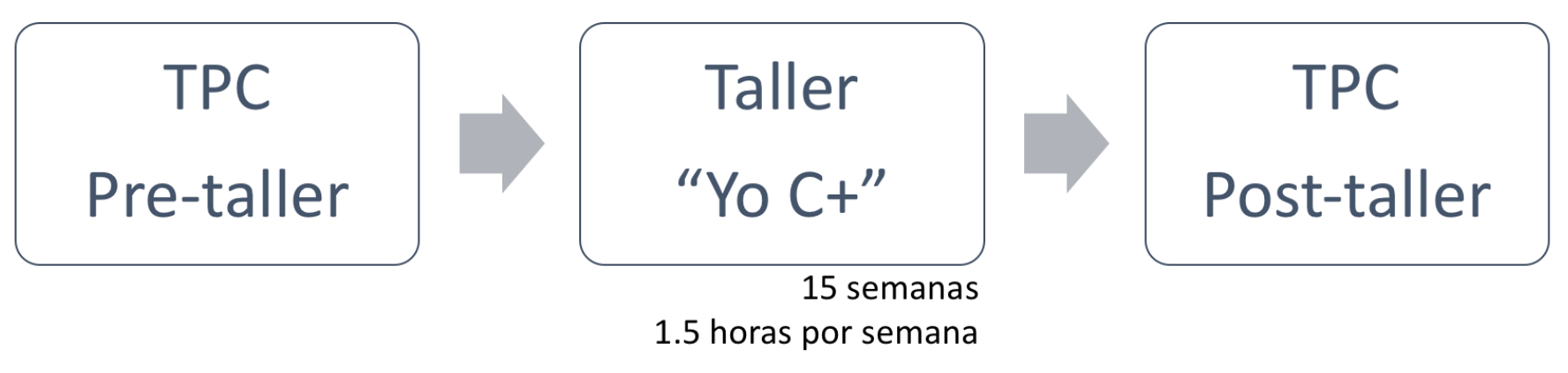}}
\caption{Diseño de la evaluación del taller}
\label{fig:pre-post}
\end{figure}

Con el fin de medir el avance del pensamiento computacional, se aplicó un Test de Pensamiento Computacional (TPC) \cite{TPC_roman} antes y después del taller (ver Figura \ref{fig:pre-post}). Este instrumento fue modificado de su versión original para ajustarlo al rango de edad de los participantes. Además, se hicieron preguntas adicionales para considerar las condiciones educacionales de quienes participaban del taller. 

\subsection{Mediciones Pre-taller y Post-taller}
El TPC \cite{TPC_roman} tiene como objetivo ``medir el nivel de aptitud y desarrollo del pensamiento computacional en el sujeto'' \cite{TPC_roman}. El constructo medido es la ``capacidad de formular y solucionar problemas apoyándose en los conceptos fundamentales de la computación, y usando la lógica y sintaxis de lenguajes informáticos de programación: secuencias básicas, ciclos, iteraciones, estructuras condicionales, funciones o eventos y variables ''\cite{TPC_roman}. 

El test se realiza \textit{online} y contiene 32 preguntas. Cada pregunta es de elección múltiple con cuatro opciones de respuesta, donde sólo una de ellas es la correcta.

El TPC permite identificar el nivel de logro alcanzado en relación a distintos conceptos computacionales, los cuales son abordados por un número de preguntas. Se consideran los siguientes conceptos: direcciones (cuatro preguntas), ciclos (ocho preguntas), condicionales (ocho preguntas) y funciones (ocho preguntas) \cite{descripcion_test_roman}.

\subsection{Taller de programación ``Yo C Más''}

El taller buscaba que los estudiantes ``desarrollen habilidades de abstracción, creatividad, algoritmos, pensamiento sistemático y resolución de problemas'', y ``que sean capaces de comprender la naturaleza científica de la disciplina'' \cite{yocmas}.  

Realizado durante el 2017, el taller incluyó 15 sesiones de dos horas enfocadas en aprender el lenguaje de programación C++. Estas se realizaban en laboratorios de computación, donde cada participante contaba con un computador con los requerimientos necesarios para que pudiera crear y ejecutar código C++ durante la sesión. Las sesiones combinaban clases expositivas e instancias de aprendizaje activo. Cada clase se dividía en dos partes. En la primera parte el instructor exponía los contenidos de la sesión, mientras que en la segunda parte cada participante debía resolver problemas usando los contenidos vistos hasta ese momento en el taller. Durante toda la clase el instructor se encargaba de entregar retroalimentación que permitiera a los participantes resolver los problemas a los cuales se enfrentaban.

La actividad fue dirigida a estudiantes secundarios que se interesaban en aprender y/o perfeccionar sus habilidades de programación. El reclutamiento de estudiantes estuvo a cargo del área de Admisión de la universidad. Los instructores de las clases fueron estudiantes de Ingeniería Civil Informática. 

El taller cubrió tópicos que incluyen el uso de variables, estructuras condicionales y de repetición junto con la creación y aplicación de estructura de datos, funciones y recursividad. Se escogió C++ como el lenguaje foco del taller debido a que es uno de los lenguajes con que se puede participar en la OCI.  

\subsection{Análisis de datos}

El estudio del nivel de pensamiento computacional al inicio del taller y su asociación a variables como el género, curso y tipo de colegio fue hecho usando regresiones lineales. Para modelar la relación de estas variables en el desarrollo del pensamiento computacional por los estudiantes que completaron el taller se utilizó \textit{ANOVA} para mediciones repetidas. 

El reporte sobre cambios en los resultados del test como en el número de respuestas correctas en los ítems computacionales: direcciones, ciclos, condicionales y funciones estuvieron sujetas a \textit{Paired T-test}. Para el análisis de las tendencias de deserción se utilizó \textit{Chi-square test for independence}.


Todos los análisis fueron realizados considerando un nivel de significancia estadística de $\alpha=0.05$ y se utilizó el \textit{software} R para toda la evaluación de resultados.

\section{Resultados}
\subsection{Participación Inicial} 

Fueron invitados a participar estudiantes de enseñanza media perteneciente a establecimientos educacionales científicos humanista de la región metropolitana. La mayor participación inicial se registró por parte de estudiantes de género masculino $(73.7\%)$, alumnos de cuarto medio $(52.5\%)$ y por estudiantes pertenecientes a colegios municipales ($51.5\%$) (Ver Tabla \ref{tabla_participacion}). 

\begin{table*}
 \caption{Participación inicial y final según variables}
\label{tabla_participacion}
\centering
\begin{tabular}{|l|c|c|c|c|c|c|c|c|c|c|}

\hline
                                      & \multicolumn{2}{c|}{\textbf{Género}} & \multicolumn{4}{c|}{\textbf{Curso de enseñanza media}} & \multicolumn{3}{c|}{\textbf{Tipo de colegio}}  & \textbf{Total} \\ \hline
\textbf{Participación según variable} & Hombre            & Mujer            & 1°           & 2°          & 3°          & 4°          & Municipal & Particular subvencionado & Privado &                \\ \hline
\textbf{Pre-taller}                   & 73                & 26               & 0            & 17          & 30          & 52          & 51        & 33                       & 15      & 99             \\ \hline
\textbf{Post-taller}                  & 24                & 6                & 0            & 10          & 11          & 9           & 15        & 8                        & 7       & 30             \\ \hline
\end{tabular}
\end{table*}
\subsection{Evaluación inicial}
De un máximo de 32 puntos, el promedio  obtenido por los 99 estudiantes que rindieron el test inicial fue de $\overline{X}=21.8$. El rango de puntajes estuvo entre los $12$ y $30$ puntos, mientras que la desviación estándar fue de $\sigma=4.05$.

Los resultados obtenidos en el test inicial fueron analizados por género, curso y tipo de colegio. Cuando se modeló los resultados obtenidos según el género del participante, una regresión lineal significativa fue encontrada. Los estudiantes de género masculino obtuvieron alrededor de $2.23$ puntos más que las mujeres ($p=0.016$). Este factor explica una varianza del $5.86\%$ de los datos. Cuando se intentó explicar los resultados según curso y tipo de colegio, ningún resultado significativo fue encontrado. 

En el caso de los conceptos computacionales, fue hallada una regresión significativa para la cantidad de respuestas correctas a las preguntas que involucraban ciclos según el tipo de colegio del participante. Los estudiantes de colegios privados obtuvieron $0.84$ puntos más que los de colegios particulares subvencionados ($p=0.028$) y $1.15$ puntos más que los de colegios municipales ($p = 0.005$) en las preguntas que abordaban este concepto. Este factor explica un $7.97\%$ de la varianza de las respuestas. En el caso de las preguntas que abordaban condicionales, fue encontrado que los hombres obtuvieron $1.43$ puntos más que las mujeres ($p=0.001$) en este ítem. El género explica un $10.5\%$ de la varianza de las respuestas. En el caso de las preguntas que involucraban direcciones y funciones no se encontraron hallazgos significativos. 

\subsection{Tendencias de deserción}

Luego de cuatro meses de realización del taller, hubo una disminución de la participación. De los 99 estudiantes que se registraron inicialmente, solo 30 de ellos finalizaron la actividad. La tasa de deserción total fue de $69.7\%$.

El género de los participantes no fue un factor relacionado de manera significativa a la deserción, $(p=0.35)$. La tasa de deserción femenina fue de $76.9\%$ y la de hombres fue de $67.1\%$.

Considerando el curso al que pertenecían los participantes, se reveló que existe una dependencia entre esta variable y las tasas de deserción, $\chi^{2}(2,N=99) = 11.28, p=0.004$. Los estudiantes de cuarto medio tienden a abandonar la actividad en mayor medida que los estudiantes de segundo y tercero medio. El grupo de alumnos de segundo medio tuvo una tasa de deserción de $41.2\%$. En el caso de los participantes que cursaban tercero medio la baja fue de $63.3\%$. Por último, hubo una baja de $82.7\%$ para el grupo que representa a los participantes de cuarto medio, los que casi en su totalidad hacen abandono de la actividad. 

También se encontró una dependencia entre tipo de establecimiento y  deserción. Los estudiantes de establecimientos particulares subvencionados tienden a abandonar en mayor medida que los participantes de  establecimientos no subvencionados (municipales y privados), $\chi^{2}(1,N=99) = 13.69, p\leq0.001$. La tasa de deserción de participantes de colegios particulares subvencionados fue de $84.3\%$,  la tasa de deserción de estudiantes provenientes de colegios municipales fue de $54.5\%$ y la de establecimientos privados fue de $53.3\%$.

\begin{figure}[!htpb]
\centerline{\includegraphics[width=\linewidth]{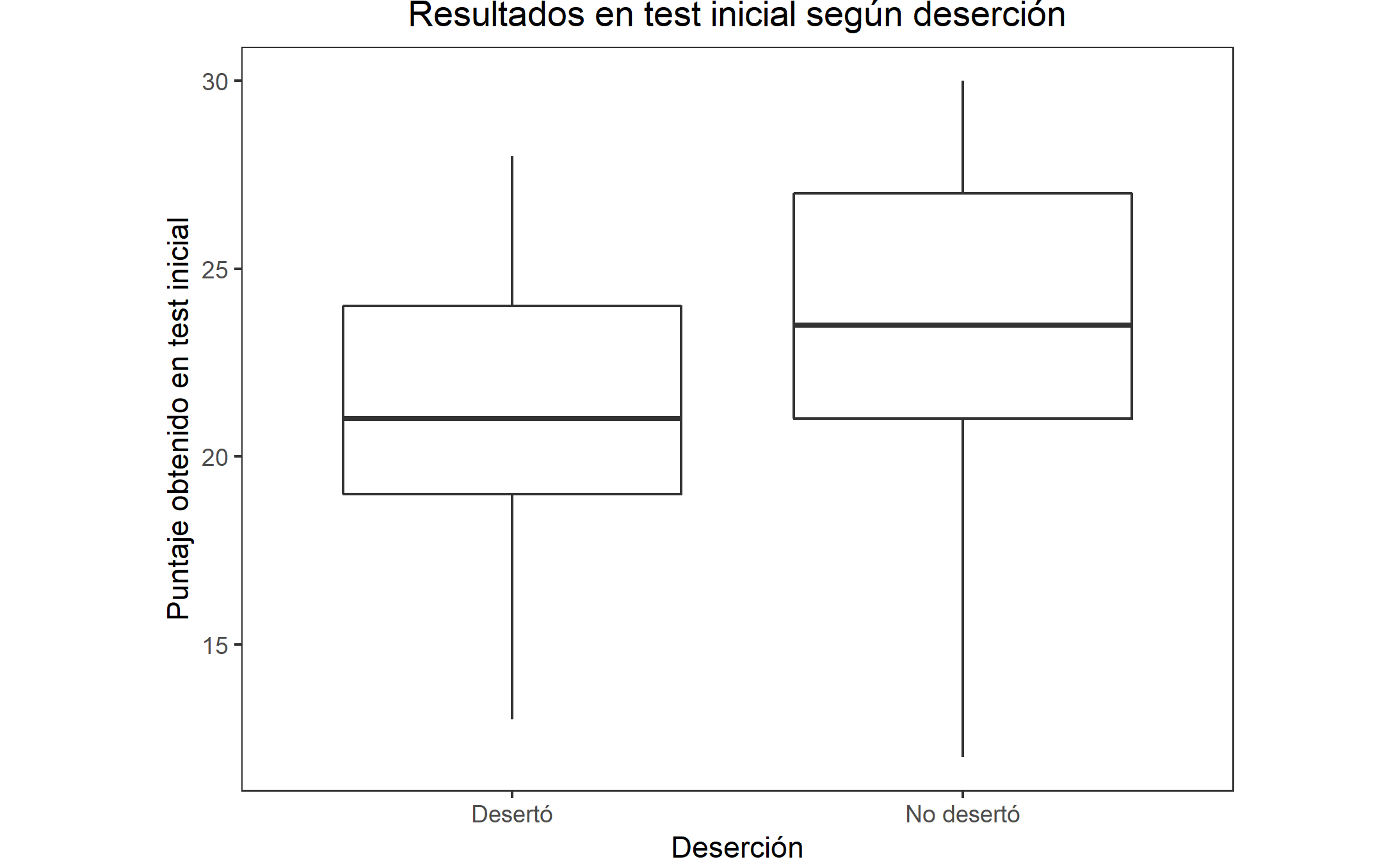}}
\caption{Resultados en test inicial según deserción}
\label{puntajeDesertoNoDeserto}
\end{figure}

Si bien 99 participantes rindieron el test de pensamiento computacional al inicio del curso, solo 30 de ellos permanecieron hasta el final. Al utilizar una regresión lineal para modelar los resultados iniciales en función de si el estudiante abandona el taller o no, se encontró que en la prueba inicial los estudiantes que desertaron obtuvieron alrededor de $2.20$ puntos menos que los estudiantes que sí completaron el taller ($p=0.013$). De hecho, cerca de un $75\%$ de los participantes que no desertaron se ubican sobre la mediana de los resultados obtenidos por estudiantes que sí abandonaron la actividad (ver Fig. \ref{puntajeDesertoNoDeserto}, ver Tabla \ref{tabla_test_inicial_segun_participacion}). 

\begin{table}[h]
\caption{Resultados en test inicial según deserción}
\label{tabla_test_inicial_segun_participacion}
\centering
\begin{tabular}{|l|c|c|}
\hline
\textbf{Resultados en test inicial}      & \textbf{Resultados en test} & \textbf{$\overline{X}$} \\ \hline
\textbf{Participantes que desertaron}    & $21.13$                     & $3.59$                  \\ \hline
\textbf{Participantes que no desertaron} & $23.33$                     & $4.73$                  \\ \hline
\end{tabular}
\end{table}

\subsection{Mejora en Pensamiento Computacional}

\begin{figure}[!htpb]
\centerline{\includegraphics[width=\linewidth]{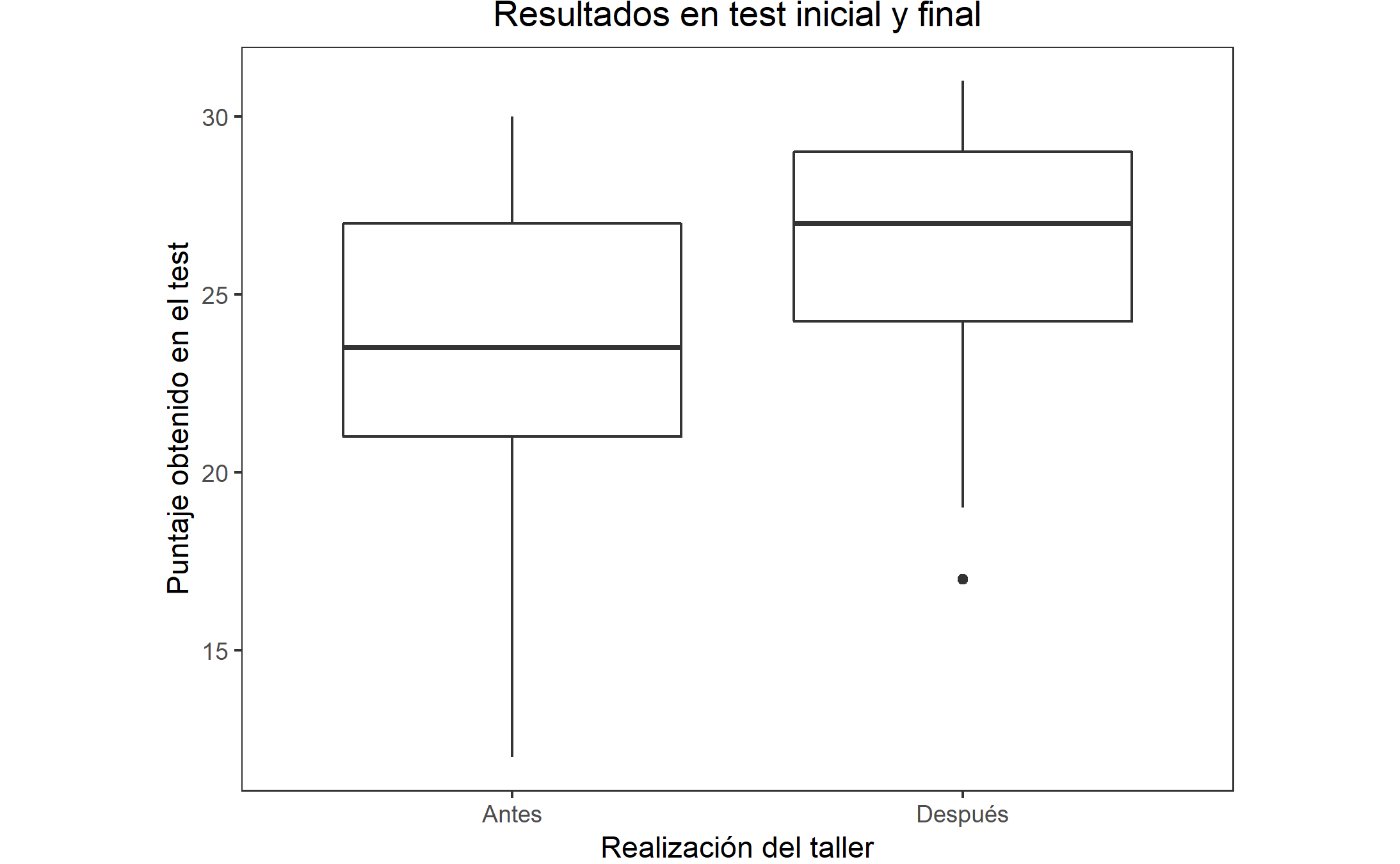}}
\caption{Resultados iniciales y finales en TPC por estudiantes que no desertaron.}
\label{avancePC}
\end{figure}

Una vez cursado el taller ``Yo C Más'', el promedio de los  puntajes obtenidos en el TPC ($\overline{X}=26.17; \sigma=3.74$) fue significativamente mayor al inicial obtenido por los mismos estudiantes ($\overline{X}=23.33; \sigma=4.73), t(29)=-4.59, p\leq0.001$ (ver Fig. \ref{avancePC}, ver Tabla \ref{tabla_avance_PC}). Estos resultados sugieren que el taller fue efectivo para desarrollar el pensamiento computacional en quienes completaron la actividad. 

\begin{table}[h]
 \caption{Resultados en test inicial y final}
\label{tabla_avance_PC}
\centering
\begin{tabular}{|l|c|c|}
\hline
\textbf{Resultados en test}           & \textbf{$\overline{X}$} & \textbf{$\sigma$} \\ \hline
\textbf{Pre-test (30 participantes)}  & $23.33$                 & $4.73$            \\ \hline
\textbf{Post-test (30 participantes)} & $26.17$                 & $3.74$            \\ \hline
\end{tabular}%
\end{table}

\begin{figure}[!htpb]
\centerline{\includegraphics[width=\linewidth]{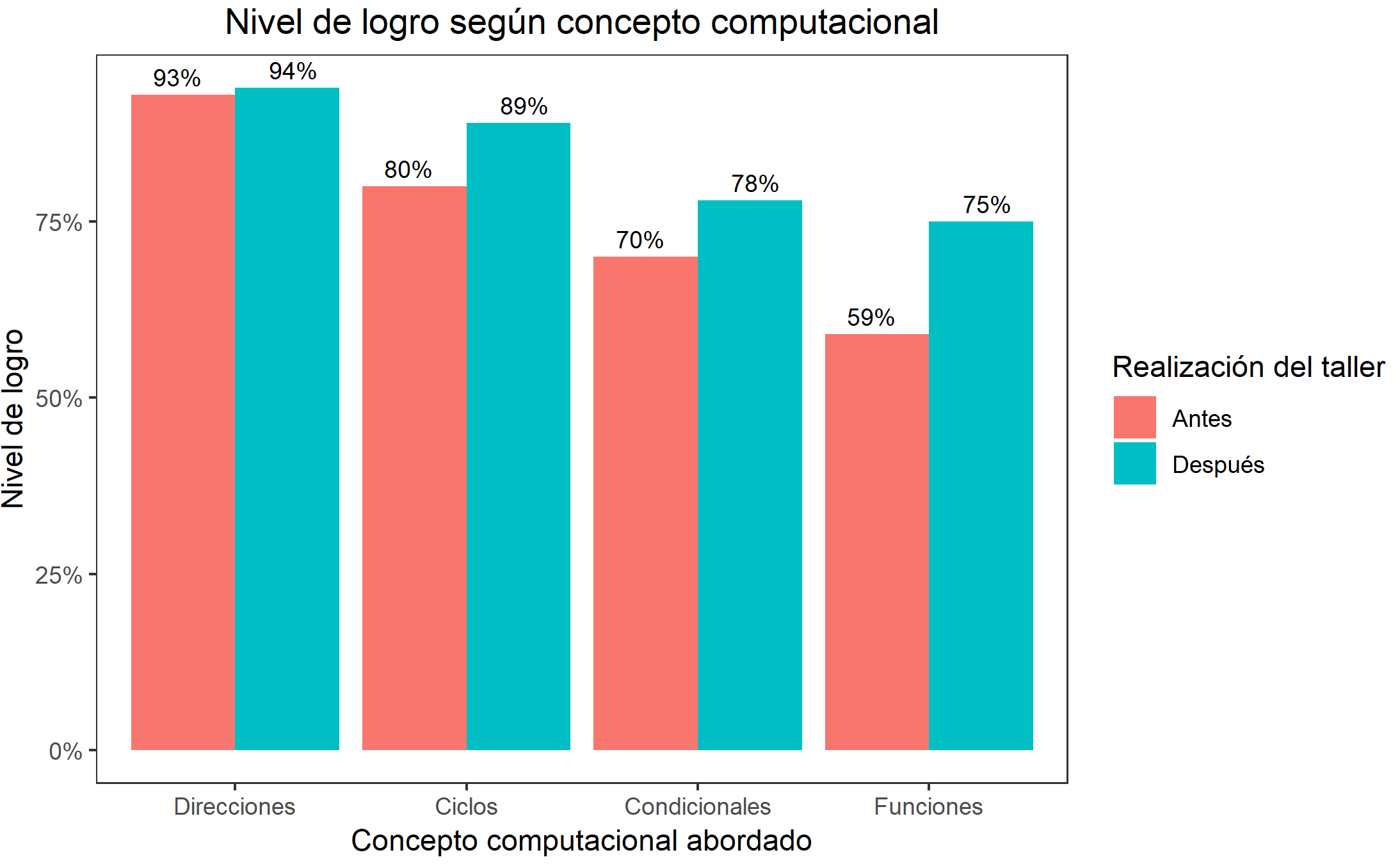}}
\caption{Nivel de logro en test inicial y final según concepto computacional abordado.}
\label{avancePCConcepto}
\end{figure}

Teniendo en cuenta que el porcentaje de logro obtenido por cada ítem (direcciones, ciclos, condicionales y funciones) está dado por el número de respuestas acertadas en esa categoría divido por el número de preguntas que la componen, en Fig. \ref{avancePCConcepto} se refleja un aumento de los porcentajes de logro en cada concepto computacional una vez finalizado el taller. Los participantes que completaron el taller mostraron un aumento en el nivel de logro del concepto direcciones en un $0.83\%$, en el caso de ciclos fue de un $8.75\%$, en condicionales aumentó $7.22\%$, por último, en el caso de funciones el número de respuestas correctas en este ítem aumentó un $15.42\%$. Fue el concepto computacional más débil inicialmente el que presentó un mayor crecimiento al finalizar el taller. 


\textit{Paired T-Test} fue aplicado para evaluar si el aumento de respuestas acertadas en cada concepto computacional fue significativo. No hubo un aumento significativo en el número de respuestas correctas del ítem de direcciones. Sin embargo, se reportó una mejora en los ítems: ciclos, condicionales y funciones. El promedio de respuestas correctas  en el ítem ciclos al finalizar el taller ($\overline{X}=7.13; \sigma = 1.07$) fue significativamente mayor al inicial ($\overline{X}=6.43; \sigma=1.41), t(29)=-2.70, p=0.006$. El número de respuestas correctas al finalizar el taller en el concepto condicionales ($\overline{X}=9.3; \sigma = 1.84$) fue significativamente mayor al inicial ($\overline{X}=8.43; \sigma=2.01), t(29)=-2.87, p=0.004$. En el ítem de funciones el número de respuestas correctas  ($\overline{X}=5.97; \sigma = 1.22$) fue significativamente mayor al inicial ($\overline{X}=4.73; \sigma=1.95), t(29)=-3.58, p\leq0.001$. Estos resultados sugieren que el taller tiene efecto en el número de respuestas correctas de los conceptos computacionales: ciclos, condicionales y funciones. Específicamente, nuestros resultados sugieren que una vez finalizado el taller el número de respuestas correctas en estos ítems aumenta.

\begin{figure*}[!htpb]
\centerline{\includegraphics[width=\linewidth]{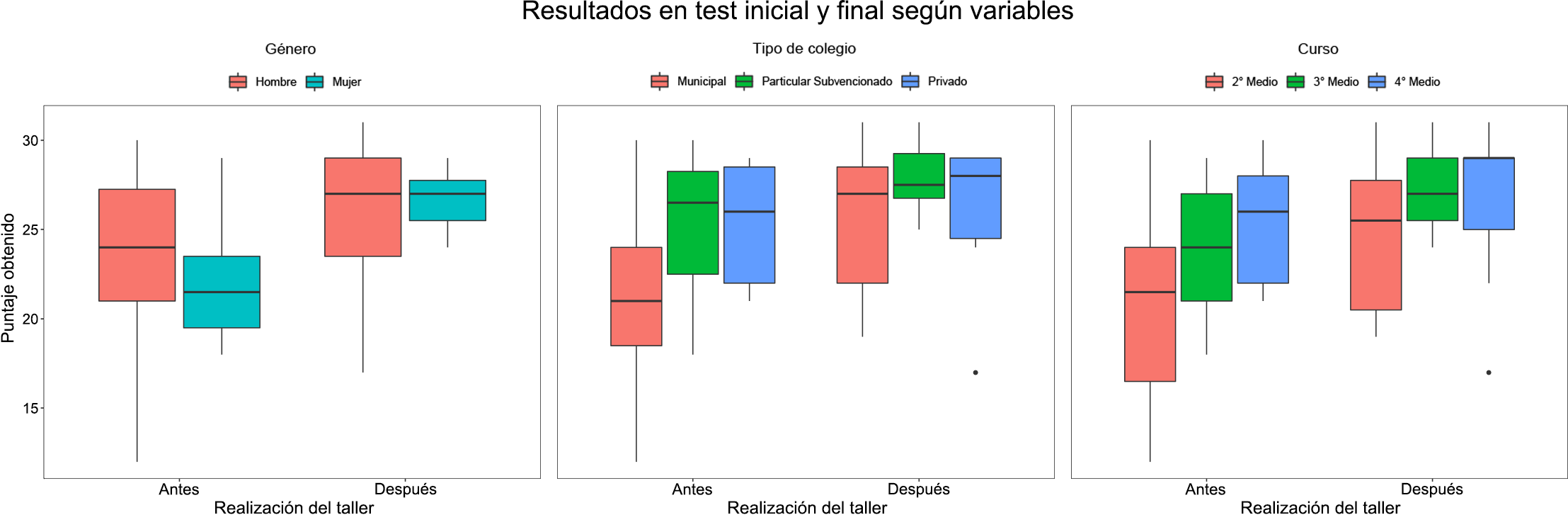}}
\caption{Resultados iniciales y finales en TPC según género, curso y tipo de colegio.}
\label{AvancePorVariable}
\end{figure*}

La Fig. \ref{AvancePorVariable} resume los resultados obtenidos por género, curso y tipo de colegio en el test inicial y final. Al analizar los resultados por género se determinó que este factor no era significativo. El análisis se realizó considerando los resultados obtenidos en el test inicial y final como también analizando la cantidad de respuestas correctas en el tiempo por cada uno de los conceptos computacionales en los que se presentó una mejora significativa: ciclos, condicionales y funciones.

Dada las pocas observaciones al finalizar el taller, para el análisis del curso y tipo de colegio los estudiantes fueron agrupados en tres categorías según la Tabla \ref{tabla_estudiantes_agrupados}:  estudiantes de segundo medio de colegios no privados (pertenecientes a colegios particulares subvencionados o municipales), estudiantes de tercero y cuarto medio de colegios no privados y estudiantes de tercer y cuarto medio de colegio privados.

\begin{table}[h]
\caption{Participación final según tipo de colegio y curso}
\label{tabla_estudiantes_agrupados}
\centering
\begin{tabular}{|l|l|l|}
\hline
\textbf{Tipo de colegio por curso} & \textbf{2° Medio} & \textbf{3° y 4° Medio} \\ \hline
\textbf{Privado}                   & 0                 & 7                      \\ \hline
\textbf{No privado}                & 10                & 13                     \\ \hline
\end{tabular}
\end{table}

\begin{figure}[!htpb]
\centerline{\includegraphics[width=\linewidth]{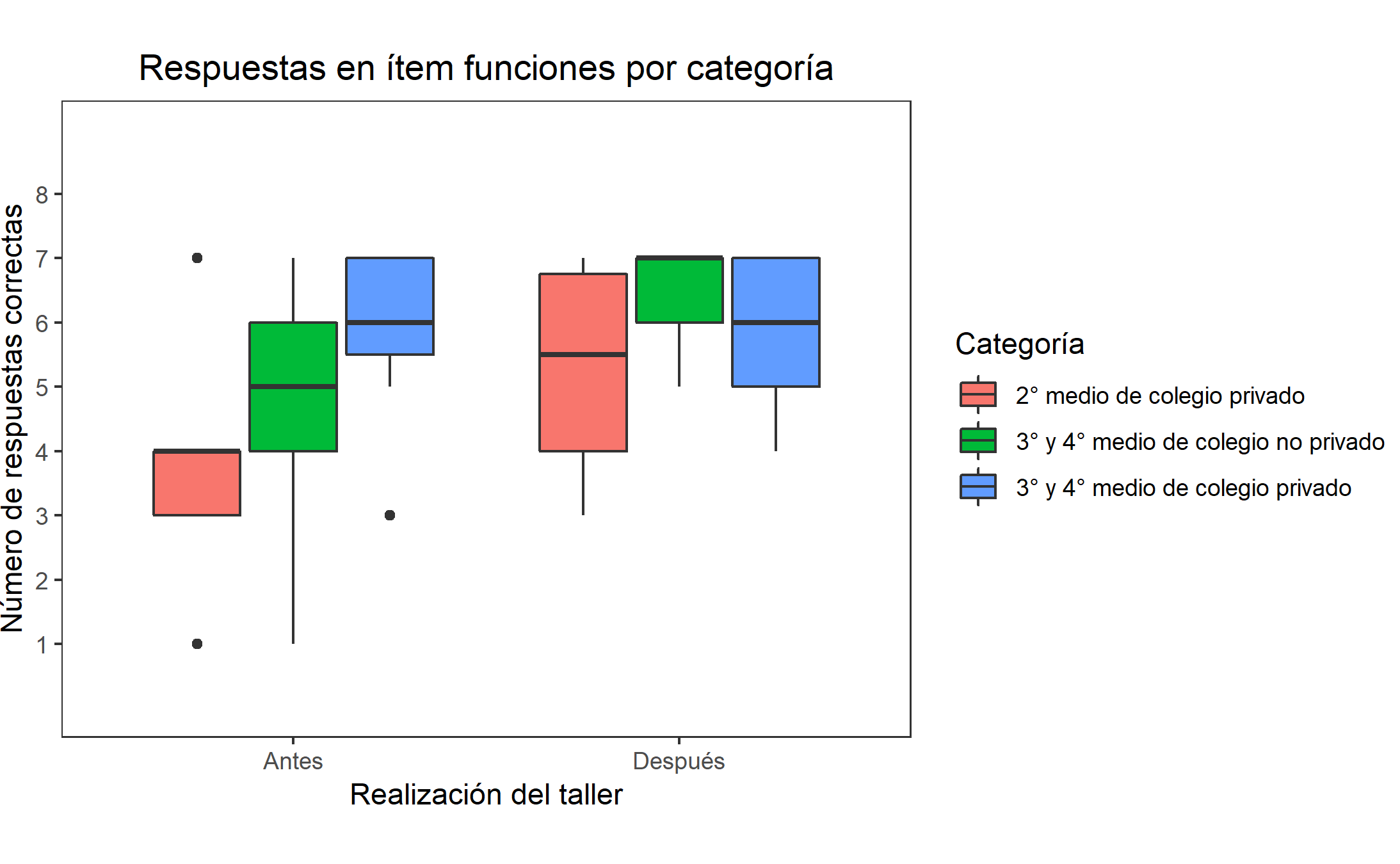}}
\caption{Número de respuestas correctas en el tiempo del ítem funciones según categoría}
\label{AvanceFuncionesNiveles}
\end{figure}

Analizando los efectos de las categorías presentadas, efectos significativos del taller fueron encontrados para el número de respuestas del ítem funciones, $F(2,27)=3.21, p =0.05$. Para este caso, una regresión lineal significativa fue encontrada. De acuerdo a ella, en general los estudiantes de tercero y cuarto medio de colegios privados tienden a  responder correctamente $1.31$ más preguntas en ese ítem  que los estudiantes de segundo medio del mismo tipo de colegio ($p=0.02$) (ver Fig \ref{AvanceFuncionesNiveles}). Comparado los primeros con respecto a los estudiantes de tercero y cuarto medio de colegios no privados no se encontraron diferencia significativas ($p=0.75$). Estas categorías junto con el tiempo explicarían una varianza del $24.06\%$ de los datos.

Si bien el tiempo siempre fue un factor significativo en los análisis. Fuera de lo señalado, no se reportaron efectos del curso o tipo de colegio en los resultados obtenidos en el test ni en la cantidad de respuestas correctas de los ítems: direcciones, ciclos y condicionales.

Por último, no se encontró evidencia de interacciones del género, curso y tipo de colegio con el factor tiempo tanto en los resultados del test como en la cantidad de respuestas acertadas en los conceptos computacionales. 

\section{Discusión}
\subsection{Evaluación y desarrollo del pensamiento computacional}



El taller realizado permitió elevar el promedio de los puntajes obtenidos en el test de pensamiento computacional ya que los resultados finales fueron significativamente mayores a los obtenidos inicialmente por los mismos estudiantes. Esta evidencia indica que un curso de programación en un lenguaje de bajo nivel puede ser un mecanismo efectivo para promover el pensamiento computacional en estudiantes de enseñanza media. 



Los resultados además indican que los participantes mostraron un aumento significativo en los porcentajes 
de logro en tres de los conceptos computacionales que abarcaba la prueba. El mayor
aumento se registró en las preguntas que abordaban funciones, seguido por ciclos y luego condicionales. Fue el concepto computacional más débil en el test inicial el que presentó una mayor tasa de aumento al finalizar el taller. Dado que en cada sesión del taller se buscaba que los estudiantes adquirieran habilidades de programación abordando distintos conceptos computacionales, las tasas de aumento reflejan la efectividad de los contenidos vistos en clases.

En el caso del curso ``Yo C Más'', el lenguaje de programación C++ fue de  utilidad. Si bien C++ se distancia de Scratch, herramienta utilizada generalmente en este tipo de experiencias orientadas a individuos sin experiencias previas de programación, los participantes del taller pudieron aumentar su dominio del pensamiento computacional utilizando un lenguaje de bajo nivel. 

Dado que en \cite{experimento_argentina} se mostró que un curso de programación en estudiantes de primer año de la Facultad de Ciencias Físico Matemáticas y Naturales de la UNSL ayudó a disminuir la deserción y mejorar el bajo rendimiento en asignaturas vinculadas a la resolución de problemas, resulta relevante que los participantes del taller ``Yo C Más'' más cercanos a ingresar a la universidad (tercero y cuarto medio) hayan sido los que mejores resultados obtuvieron al finalizar el taller. Pareciera ser que orientar talleres de programación con un lenguaje de bajo nivel a sujetos de estos cursos sería un buen método para fomentar el pensamiento computacional y podría proveer beneficios en el rendimiento académico de quienes ingresen a la universidad.


\subsection{Tendencias en los participantes}


Hubo mayor participación de estudiantes de género masculino, alumnos de cuarto medio y por estudiantes de colegios municipales al inicio del proyecto. Teniendo en consideración que el llamado a participar de la actividad estuvo a cargo del Departamento de Admisión de la universidad en donde se realizó el taller, los resultados están sujetos a quienes son el público objetivo en la promoción de este tipo de actividades y revelan las características de los que aceptan este tipo de invitaciones.

Se desconocen las causas para la alta tasa de deserción ($69.7\%$), parte de ella puede ser explicada por la discontinuidad de las sesiones realizadas debido a problemas en la disponibilidad de los laboratorios en los que se realizaba el taller junto con que el taller era de carácter voluntario y gratuito. La alta tasa de deserción por estudiantes de cuarto  medio ($82.7\%$) puede estar asociada a la falta de tiempo de éstos para participar en este tipo de actividades dado que están estudiando para las pruebas de selección universitaria. Con respecto a la alta tasa de deserción de estudiantes de colegios particulares subvencionados ($84.3\%$), no se tienen hipótesis sobre cuáles pueden ser las razones. Estudios futuros debieran investigar las causas de abandono de los participantes en este tipo de actividades.

Además, los resultados del test inicial revelaron que quienes abandonaron tuvieron peores resultados que los estudiantes que sí completaron el taller. Esto indica que las motivaciones y nivel de compromiso de los estudiantes con esta actividad puede estar relacionada con cierto nivel inicial de habilidades vinculadas al pensamiento computacional.

\subsection{Limitaciones}

Como todo estudio empírico, nuestro estudio presenta limitaciones. A pesar que el test utilizado ha sido validado \cite{TPC_roman}, sus propias limitaciones han sido reportadas \cite{modificacion_curriculum}. Si bien el test considera fuertemente la dimensión del pensamiento computacional: conceptos computacionales, abarca parcialmente  prácticas computacionales e ignora perspectivas computacionales. 

Además, dado que la selección de los participantes del estudio no fue aleatoria, el diseño quasi-experimental pudiese no considerar todos los factores preexistentes (por ejemplo, grado de conocimientos previos de programación adquiridos en sus establecimientos educacionales) ni las influencias externas que pudieron haber afectado los resultados. A pesar de lo anterior, entendiendo los resultados  dentro de su contexto y considerando que la muestra utilizada para medir el desarrollo del pensamiento computacional fue pequeña (30 estudiantes), lo obtenido permite obtener una evidencia previa a experimentos cuantitativos centrados en las razones subyacentes de los resultados generados.





\section{Conclusiones}
A través de un caso de estudio, este documento entrega evidencias del impacto de un taller de programación que utiliza un lenguaje de programación de bajo nivel (C++) en estudiantes de enseñanza media pertenecientes a colegios científicos humanistas de Chile. La aplicación de un test de pensamiento computacional al inicio y final del taller permitió revelar que una vez finalizado la actividad hubo una mejora significativa de los resultados obtenidos en el test. En particular en el número de respuestas correctas en los conceptos computacionales: ciclos, condicionales y funciones. 
Sin embargo, dado de que en el ítem funciones los mejores resultados son obtenidos por estudiantes de tercero y cuarto medio, los resultados sugieren que un taller de programación basado en un lenguaje de bajo nivel es más efectivo en sujetos de cursos superiores.

Considerando las altas tasas de deserción, trabajos futuros debieran determinar cuáles son las causas de abandono de los participantes y se debiesen emplear medidas para aumentar la tasa de retención. Por último, se debe investigar si la motivación y nivel de compromiso en este tipo de actividades está sujeto a un cierto nivel inicial de habilidades vinculadas al pensamiento computacional. 

\section*{Agradecimientos}
Agradecemos a Marcos Román-González, profesor de la Universidad Nacional de Educación a Distancia, por facilitar una versión del test de pensamiento computacional modificada al rango de edad de los participantes del estudio. Además, agradecemos a Pedro Godoy, junto a los tutores y participantes del taller ``Yo C Más'', quienes permitieron realizar esta investigación. La segunda autora también agradece el apoyo otorgado por CONICYT, bajo el proyecto FONDECYT Iniciación \#11161026.

\bibliography{referencias}

\bibliographystyle{IEEEtran}

\end{document}